\documentclass[]{iopart}
\usepackage{graphicx}
\usepackage{wrapfig}

\newcommand{\bra}[1]{\langle#1|}
\newcommand{\ket}[1]{\left|#1\right\rangle}

\newcommand{\ketb}[1]{|#1\}}
\newcommand{\brab}[1]{\{#1|}
\newcommand{\ketp}[1]{|#1)}

\newcommand{\foster}{F\"{o}ster\ }

\begin{document}
 \title{Zeros of Rydberg-Rydberg \foster Interactions}

\author{Thad G. Walker and Mark Saffman}
\address{Department of Physics, University of Wisconsin-Madison,
		Madison, Wisconsin 53706 }
\date{\today}


\begin{abstract} 
Rydberg states of atoms are of great current interest for quantum man\-ip\-ulation of mesoscopic samples of atoms.  Long-range
Rydberg-Rydberg interactions can inhibit multiple excitations of atoms under the appropriate conditions.  These
interactions are strongest when resonant collisional processes give rise to long-range $C_3/R^3$ interactions.  We show in this
paper that even under resonant conditions $C_3$ often vanishes so that care is required to realize full dipole blockade in
micron-sized atom samples.  
\end{abstract}

\ead{tgwalker@wisc.edu}

\section{Introduction}
Rydberg-Rydberg interactions are very interesting for use in mesoscopic quantum manipulations.  The extremely strong interactions
between two Rydberg atoms have been proposed to entangle large numbers of atoms in a non-trivial manner using the phenomenon of
 blockade \cite{Lukin01}.  When a mesoscopic sample is illuminated with narrowband lasers tuned to a Rydberg state, only
one atom at a time can be excited if the Rydberg-Rydberg interaction exceeds the linewidth of the laser-atom coupling. 
Therefore the mesoscopic cloud of atoms behaves as an effective 2-level system, with the upper level being a single collective
excitation.

In order to attain the strongest possible Rydberg blockade, it is desirable to operate under conditions where the Rydberg-Rydberg
interaction is not the usual $C_5/R^5$ or $C_6/R^6$ van-der-Waals interactions, but instead is  resonantly enhanced
by ``\foster$\!$'' processes \cite{Lukin01,Foster96} such as \begin{equation}
ns+ns\rightarrow np+(n-1)p
\end{equation} that lead to isotropic $C_3/R^3$
long-range behavior when the $ns+ns$ states are degenerate with the $np+(n-1)p$ states.  Dramatic enhancements of collisional
interactions due to such resonant couplings have been demonstrated for Rydberg excitation in a magneto-optical
trap\cite{Fioretti99}. The quantum nature of these types of interactions was recently used to resolve molecules 12 nm apart in
an organic solid \cite{Hettich02}. Due to the high density of Rydberg states, there are typically many candidate levels for such
\foster processes.

An important consideration for quantum information applications of Rydberg blockade is the size of cloud required.  The spatial
extent of the cloud must be small enough for excitation of a single atom anywhere in the cloud to block the excitation of every
other atom in the cloud.  Even with the great strength of Rydberg-Rydberg interactions, the required size of the clouds is likely
to be less than 10 microns for fast (1 MHz) operations.  Only recently have atom clouds of such small size been produced in the
laboratory\cite{Peil03,Sebby04}.  Even in the case that sub-micron mesoscopic samples are realized experimentally, there are
other applications of such samples that benefit from the sample being as large as possible.  For example, we have recently
proposed single-photon sources, single-photon detectors, and quantum state transmission \cite{Saffman02,Saffman04} using
mesoscopic Rydberg blockade.  In these applications one would like the cloud to be many wavelengths in size in order that the
diffraction limited light fields not occupy too large a solid angle.  For this to be effective requires the blockade to operate 
over clouds of several microns in extent.

The purpose of this paper is to examine some issues that arise in the application of \foster processes to Rydberg blockade.  In
particular, when the quantum numbers for the states are of the form
\begin{equation}
nl+nl\rightarrow n'(l-1)+n''(l\pm 1)
\end{equation}
we show that one or more of the $nl+nl$ states have $C_3=0$.  These ``\foster zero'' states then only exhibit the usual
van-der-Waals long-range behavior and will therefore set limits on the attainable cloud sizes for quantum manipulations
using Rydberg blockade. Only when $l'=l''=l+1$ are there no \foster zero states. Recent experiments \cite{Tong04,Singer04} have
observed strong suppression of excitation in very large clouds that is strong evidence for blockade but do not address whether
or not the blockade is complete as required for quantum information applications.

In the following, we first (Section \ref{sec:background}) present a more detailed background discussion of the importance of
strong isotropic Rydberg-Rydberg interactions for quantum manipulations using Rydberg blockade.  We then illustrate in Section
\ref{sec:foster} how the
\foster process accomplishes this.  Section \ref{sec:zeros} presents the main result of this paper: for many possible \foster
processes there exists a linear combination of atomic sublevels that have $C_3=0$.  This result is extended to the important
case of fine-structure interactions in Section \ref{sec:finestructure}.  We conclude with more implications of this work.

\section{Background}\label{sec:background}

\begin{figure}[ht]
\centering\includegraphics[scale=0.75]{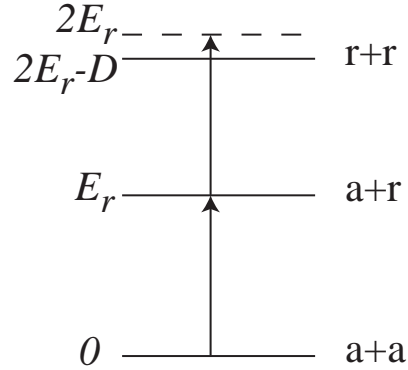}\hspace*{0.5 in}
\caption{Energy levels for a pair of atoms being excited by a light field to Rydberg states.  Excitation of one
of the two atoms is resonantly favored but two-atom excitations are off resonance by the dipole-dipole
interaction energy $D$.}\label{entangle}
\end{figure}

The basic physics behind the dipole-blockade
concept is illustrated in Figure~\ref{entangle}.  Suppose we have $2$ atoms in an atomic ground state $a$.  We envision exciting
the atoms to a Rydberg state $r$ of energy $E_r$ using a narrowband laser.  As shown in the figure, excitation of one of the two
atoms to the Rydberg state is allowed while excitation of the second atom is off-resonant and therefore suppressed.  Addition of
more atoms changes the effectiveness of this ``blockade'' but does not change the basic physics of the situation. 
Excitation of more than one atom is energetically suppressed as long as the interaction between each pair of atoms
in the ensemble exceeds the bandwidth of the laser-atom excitation.  Neglecting spontaneous emission or other decohering
effects, when subject to dipole blockade and a continuous light field the atoms will undergo coherent Rabi oscillations between
states
$a+a$ and $a+r$.

To see how this process can be used to generate interesting entanglement, we simply imagine driving the atom pair in the
$\ket{aa}$ state with a
$\pi$ pulse.  The atoms will then be in a coherent superposition of states $a$ and $r$
\begin{equation}
\psi(\pi)={-i\over \sqrt{ 2}}\left(\ket{ar}+\ket{ra}\right)
\end{equation}
that cannot be written as a product of two individual wavefunctions and is therefore entangled.  To avoid the inevitable
decoherence of the unstable Rydberg state, the entanglement can be usefully transferred to a second ground state $b$ with a
$\pi$-pulse from a second laser tuned to the $r\rightarrow b$ transition.  Again, this concept can be extended to a collection
of $N$ atoms without loss of generality.  The final wavefunction is a symmetric entangled superposition of $N-1$ atoms in state
$a$ and one atom in state $b$.

There are several intriguing characteristics of this process.  First, the entanglement is generated between the internal states
of the atom; any motion of the atom is unimportant.  This means that it is not necessary for the atoms to be localized in the
ground state of a confining potential, so the method does not require coherence of the external degrees of freedom of the
atoms.  There is also no motional constraint on the speed at which the process can occur. Second, the value of the
Rydberg-Rydberg interaction
$D(R)$ is not important to first order, as long as it is much larger than the bandwidth of the light-atom interaction.  This
implies that the atoms can be at random distances
$R$ from each other, as  long as they are not too far apart.  Finally, since the blockade mechanism suppresses excitation of
multiple Rydberg atoms, the atoms never actually experience the strong Rydberg-Rydberg forces, avoiding heating.

Key to the entanglement process is the requirement that the Rydberg-Rydberg frequency shift be large compared to the bandwidth
of the light-atom interaction.  If this is violated, the fidelity of the entanglement operations will be compromised. In a
mesoscopic sample, insufficient blockade in any possible excitation channel is sufficient to cause copious production of excited
atoms. We now examine the Rydberg-Rydberg interactions to see under what conditions this will be problematic.

\section{\foster Process}\label{sec:foster}

\begin{figure}
\centering\includegraphics[scale=0.8]{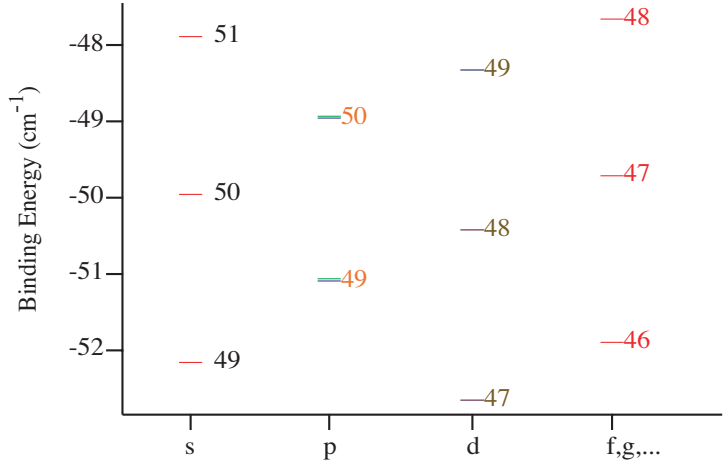}
\caption{Left: Energy levels  for Rb near $n=50$. }\label{Rbelevels}
\end{figure}

We wish to consider the interactions between two like Rydberg atoms that lead to blockade.  Normally, in the absence of
an external electric field these interactions are of the van-der-Waals forms $1/R^5$ or $1/R^6$. Since an electric field mixes
states of opposite parity together,  a Rydberg atom can have a permanent dipole moment, leading to a much stronger and
longer range $1/R^3$ interaction. The interaction between two such atoms A and B is the familiar
\begin{equation}
V_{DD}={1\over R^3}(3{\bf p}_A\cdot \hat{\bf R}\hat{\bf R}\cdot {\bf p}_B-{\bf p}_A\cdot {\bf p}_B)={p^2\over
R^3}(3\cos^2\theta-1)
\end{equation}
where $\theta$ is the angle between the interatomic separation $\bf R$ and the electric field ${ E}\bf\hat{z}$. The fact that
$p\sim n^2e a_0$ makes this interaction huge for two Rydberg states.  However, it has the undesirable feature for
dipole-blockade that it vanishes at
$\theta=\cos^{-1}\sqrt{1/3}=54.7^\circ$, allowing for excitation of Rydberg atom pairs located at this angle with respect to
each other.

It is possible to have an isotropic Rydberg atom-atom interaction of comparable strength to this if there is a degeneracy in the
energy-level spectrum for a pair of atoms.  For example, inspection of Figure~\ref{Rbelevels} shows that the 50s
state of Rb is nearly symmetrically placed between the 49p and 50p states.  This means that the 50s+50s state of a Rydberg atom
pair is nearly degenerate with the 49p+50p state. Neglecting the influence of other nearby
states, the eigenstates of the two-atom system are linear combinations of 50s+50s and 49p+50p, with energy shifts proportional
to $1/R^3$.   Using the methods described below, we find that the
Rydberg-Rydberg potential energy curves are given by
\begin{equation}
U_\pm(R)=\frac{\delta }{2} \pm {\sqrt{\frac{4U_3(R)^2}{3} + \frac{{\delta }^2}{4}}}
\end{equation}
where $U_3(R)=e^2\langle50s||r||50p\rangle\langle50s||r||49p\rangle/R^3=5.75\times10^3$ MHz $\mu$m$^3/R^3$ and the ss-pp energy
defect is $\delta=E(49p)+E(50p)-2E(50s)=-3000$ MHz.  Estimates similar to this have been previously presented by Protsenko {\it
et al.}\cite{Protsenko02}.

\begin{figure}
\centering\includegraphics[scale=0.6]{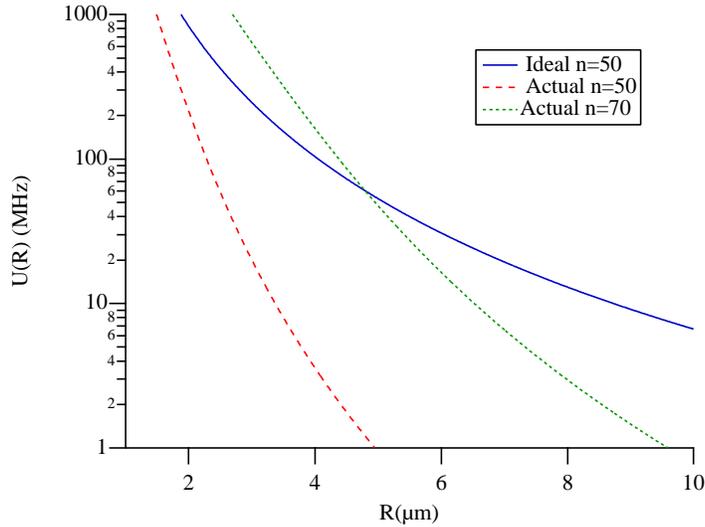}
\caption{Isotropic dipole-dipole interaction for excitation to the Rb 50s state.  The energy defect between the 50s+50s and
49p+50p states significantly reduces the interaction as compared to the ideal degenerate case. Going to larger $n$
partially compensates for the non-zero energy defect.}\label{pot50s}
\end{figure}

A plot of $U_+(R)$ is shown in Figure~\ref{pot50s}.  If there were perfect \foster degeneracy ($\delta=0$) the dipole-dipole
interaction would exceed 10 MHz out to distances of 10 microns.  This would be very promising for realizing dipole blockade in
quite large clouds.  However, the energy defect for real Rb atoms reduces the interaction substantially, becoming the usual
van-der-Waals form
$-4U_3(R)^2/3\delta $ at large distances.  Here we should note that we are restricting ourselves to mixing of states that are
nearly resonant.  Other states, such as 51p+48p in this example, have larger energy defects ($\delta=12$ GHz for 51p+48p)  and
often much smaller matrix elements as well.  Adding up many such states could (and should) be done in second order perturbation
theory and will alter the
$1/R^6$ part of the potential somewhat but not change the overall conclusions.  For the rest of this paper we will continue to
neglect these other states.  We will also neglect the $1/R^5$ quadrupole-quadrupole interactions.

The strength of the Rydberg-Rydberg interactions varies strongly with principal quantum number.  The radial matrix elements
scale as $n^2$, so $U_3$ scales as $n^4$.  Since the energy defect $\delta$ scales as $n^{-3}$, the van-der-Waals interactions
scale as $U_3^2/\delta\sim n^{11}$.  Thus the range of the Rydberg blockade increases as roughly $n^{11/6}$, so that $n=70$ has
nearly twice the range as $n=50$, as illustrated in Figure~\ref{pot50s}. Increasing $n$ comes with increased sensitivity to stray
electric fields, black-body radiation, and so forth.  In practice, states up to $n\sim 100$ should be stable enough to perform
fast quantum operations\cite{Ryabtsev04}. However, it would be preferable to find states with smaller energy defects and
therefore larger range at smaller
$n$.

Given the lack of perfect \foster resonances for the s-state, one should consider the possibility of
\foster processes in other angular momentum states where the energy defects may be smaller than they are for the $s$ states.

\section{\foster zeros}\label{sec:zeros}

Interesting new properties of the \foster process arise when we consider higher angular momentum states.  For example, 
consider again the s and p states of Rb, but this time with laser excitation of the p-states, so that the relevant \foster
degeneracy is, say, 50p+50p$\rightarrow$50s+51s with an energy defect $\delta=$930 MHz. Suppose also that  the
atoms are subject to linearly polarized light polarized along the laboratory $\hat{\bf z}$-axis.  We shall show that there is no
\foster blockade for this situation.  For simplicity, let us assume that two atoms are aligned with $\hat{\bf R}=\hat{\bf z}$. 
In this case, consider the wavefunction
\begin{equation}
\ket{\psi_0}={1\over \sqrt{3}}\ket{50p1\;50p\bar{1}}-{1\over \sqrt{3}}\ket{50p0\;50p0}+{1\over
\sqrt{3}}\ket{50p\bar{1}\;50p1}\label{zeroppss}
\end{equation} 
($\bar{1}=-1$) whose matrix element of $V_{DD}$ is zero with the $s+s$ states:
\begin{equation}
\bra{50s\;51s}V_{DD}\ket{\psi_0}=0 \label{zeroppss2}
\end{equation}
and so the only long-range interaction will be a van-der-Waals interaction with the comparatively far off-resonant $d+d$ and
$d+s$ states.  Note that $\ket{\psi_0}$ is strongly coupled to the light through its $ \ket{50p0\;50p0}$ part.  With strong light
coupling and weak dipole-dipole coupling, we conclude that the Rb p-states will not experience long-range dipole blockade. 
These conclusions are not changed when $\hat{\bf R}$ is rotated away from $\hat{\bf z}$.  If one takes the quantization axis
along $\hat{\bf R}$ then $\ket{\psi_0}$ stays of the same form, but each of the three parts of $\ket{\psi_0}$ will contribute to
the light-atom interaction.  For the rest of the paper we shall take the quantization axis for the atomic basis states to be
along
$\hat{\bf R}$.

Another very interesting possibility from Figure~\ref{Rbelevels} is the nearly degenerate combination 48d+48d$\rightarrow$50p+46f
with an energy defect of only $110$ MHz (neglecting fine structure for now).  This has the potential for much stronger \foster
interactions at large distance as compared to the $s+s$ states.  In this case there is also a wavefunction with zero coupling via
$V_{DD}$ to the p+f manifold:
\begin{equation}
\ket{\psi_0}=\frac{1}{{\sqrt{107}}}\ketb{48d0\;
48d0}+{\sqrt{\frac{8}{107}}}\ketb{48d1\;48d\bar{1}}+
  {\sqrt{\frac{98}{107}}}\ketb{48d2\;48d \bar{2}}\label{zeroddpf}
\end{equation}
where interchange-symmetric kets are defined in terms of the quantum numbers $A=n_Al_Am_A$ of the individual atoms  as
$\ketb{A B}=(\ket{A B}+\ket{B A})/\sqrt{2+2\delta_{m_Am_B}}$, and $\bar{m}=-m$.  We shall label such states as \foster zero
states.

Whereas the \foster zero state of Equation~\ref{zeroppss} can be deduced in a straightforward way essentially by inspection once the
matrix elements of s+s with the three p+p combinations are calculated, the d+d \foster zero state of Equation~\ref{zeroddpf} is more
subtle since its matrix elements with each of the three $\ketb{pm\;f\bar{m}}$ states must vanish. Thus we now discuss in
more generality the conditions for \foster zero states to exist for the process
\begin{equation}
nl_0+nl_0\rightarrow n'l_1+n''l_2
\end{equation}
We will assume without loss of generality that $l_1\le l_2$.

The \foster zero state, if it exists, is
written as a linear combination $\ket{\psi_0}=\sum_{m_0=0}^{l_0}c(m_0)\ketb{l_0m_0\;l_0\bar{m}_0}$, so the condition
$V_{DD}\psi_0=0$ gives
\begin{equation}
\sum_{m_0=0}^{l_0}c(m_0)\brab{l_1m\;l_2\bar{m}}V_{DD}\ketb{l_0m_0\;l_0\bar{m}_0}=0\label{eqzeros}
\end{equation}
which is effectively a generalization of Equation~\ref{zeroppss2} to the case where there are multiple possible final states.
There are three cases of interest. For $l_1=l_0-1$ and $l_2=l_0+1$ (of which d+d$\rightarrow$p+f is an example) there are
$2l_1+1=2l_0-1$ equations in the $l_0+1$ unknowns $c(m_0)$. But the reflection symmetry
$\brab{l_0m_0\;l_0\bar{m}_0}V_{DD}\ketb{l_1m\;l_2\bar{m}}=\brab{l_0m_0\;l_0\bar{m}_0}V_{DD}\ketb{l_1\bar{m}\;l_2m}
$  means that $2l_1$ of the equations are the same, leaving $l_0$ equations in $l_0+1$ unknowns and therefore a
solution exists. For the case $l_1=l_2=l_0-1$ the same argument holds.  The final case, with $l_1=l_2=l_0+1$, has
Equation~\ref{eqzeros} with
$l_0+2$ equations in $l_0+1$ unknowns and therefore no \foster zero state.

These results can also be understood from the point of view of the molecular symmetries of the problem.  Following the analysis
of Marinescu\cite{Marinescu97}, the
\foster zero states have the molecular symmetries $^3\Sigma_u^+$ and $^1\Sigma_g^+$ depending on their triplet or singlet spin
character.  For
$l_1\ne l_2$, the orbital exchange-symmetric kets as defined above are not eigenstates of the operator $\sigma_v$ that reflects
the wavefunction about a plane containing the interatomic axis.  Taking into account this symmetry, we define kets
\begin{equation}
\ketp{l_1ml_2\bar{m}}=\ket{l_1ml_2\bar{m}}+\beta\ket{l_2\bar{m}l_1m}+\lambda\ket{l_1\bar{m}l_2{m}}+
\lambda\beta\ket{l_2{m}l_1\bar{m}}
\end{equation}
which have molecular parity  ($g,u$) if $p=(\bar{1})^{l_1+l_2+S}\beta=(1,\bar{1}$) and reflection symmetry eigenvalues
$\sigma=\beta\lambda$, giving rise to states of traditional molecular designation $^{2S+1}\Sigma_{g,u}^\sigma$.  Considering
triplet states, and assuming $l_1<l_2$, there are $l_1+1$ $\lambda=\beta=1$ $^3\Sigma_{u}^+$ states, $l_1+1$
$\lambda=1,\beta=\bar{1}$
$^3\Sigma_{g}^-$ states, $l_1$ $\lambda=\bar{1},\beta=1$ $^3\Sigma_{u}^-$ states, and $l_1$
$\lambda=\beta=\bar{1}$
$^3\Sigma_{g}^+$ states.  Thus for the $d+d\rightarrow p+f$ problem, there are three $^3\Sigma_u^+$ states with $d+d$
character but only two of $p+f$ character.  Therefore it is always possible to find a linear combination of the three
$^3\Sigma_u^+$ 
$d+d$ states that has zero coupling to the two $p+f$ $^3\Sigma_u^+$  states.  The reasoning is identical for the  $^1\Sigma_g^+$
symmetry.

 To summarize, the isotropic $C_3/R^3$ dipole-dipole interaction generated by the \foster process $l_0+l_0\rightarrow l_1+l_2$
will have states with $C_3=0$ unless $l_1=l_2=l_0+1$.

The above analysis has emphasized the $^3\Sigma_u^+$  and  $^1\Sigma_g^+$ states that are degenerate in the absence of
overlap.  Similar reasoning shows that for the case $l_1=l_0-1,l_2=l_0+1$ the $^3\Sigma_g^-$ and $^1\Sigma_u^-$ states have no
\foster zeros, nor do states of
$\Lambda>0$.  In general, all of these states are coupled to the light field, but  the $^3\Sigma_u^+$  and 
$^1\Sigma_g^+$ states destroy the complete blockade.

\section{Fine Structure Effects} \label{sec:finestructure}

In general, the fine stucture interaction cannot be neglected.  At $n=50$ the Rb p-state fine structure splitting is about 800
MHz and the d splitting is 100 MHz, so that at micron-scale distances there will be strong fine-structure mixing by the
dipole-dipole interaction.  At long enough range, where the dipole-dipole interaction is smaller than the fine-structure
splitting, we can use the same type of arguments as above to analyze the problem.

Let us consider the \foster process
\begin{equation}
l_0j_0+l_0j_0\rightarrow l_1j_1+l_2j_2
\end{equation}
We are mostly interested in states with total $m_j=0$.  As before, the $l_0j_0+l_0j_0$ states are symmetric linear combinations 
$\ketb{l_0j_0m\;l_0j_0\bar{m}}$.  For alkali atoms with half-integer $j$ there are $j+1/2$ such states.  On the other hand, there
are (assuming $j_1\le j_2$) $2j_1+1$ $\ketb{l_1j_1m_0\;l_2j_2\bar{m}_0}$ states. Half of these are removed from consideration
due to the
$$\brab{l_0j_0m_0\;l_0j_0\bar{m}_0}V_{DD}\ketb{l_1j_1m\;l_2j_2\bar{m}}=
\brab{l_0j_0m_0\;l_0j_0\bar{m}_0}V_{DD}\ketb{l_1j_1\bar{m}\;l_2j_2m}
$$
symmetry.  Thus the system of equations for the \foster zero amplitudes $c(m)$ has $j+1/2$ equations in $j_1+1/2$ unknowns.
A normalizable solution will exist only for $j_1<j$.
It follows that potential \foster processes such as $nd_{3/2}+nd_{3/2}\rightarrow(n+2)p_{1/2}+(n-2)f_{5/2}$ will have zeros.

\begin{figure}
\centering\includegraphics[scale=0.6]{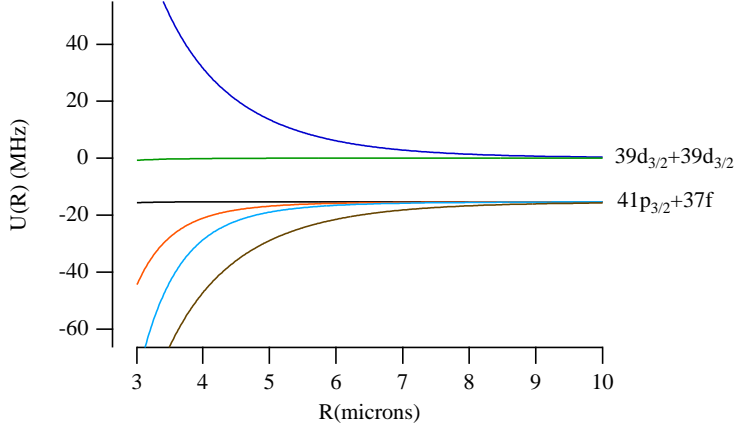}
\caption{Long-range $0_{gu}^{\pm}$ potential curves near the $39d+39d$ asymptote of Rb$_2$.  The nearly flat curve coming from
the
$39d+39d$ asymptote implies suppressed dipole blockade. }\label{fig:finestructure}
\end{figure}

The inclusion of fine structure brings new possibilities, however, since states of different $l$ can have the same value of
$j$.  For example,
\begin{equation}
nd_{3/2}+nd_{3/2}\rightarrow(n+2)p_{3/2}+(n-2)f_{5/2}
\end{equation}
 has an energy defect of only -15 MHz at $n=39$ and should not have any \foster zeros.  The Hamiltonian matrix has the
structure
\begin{equation}
H=\left(\begin{array}{cc}
{\rm diag}(0)&W U_3(R)\\W U_3(R)&{\rm diag}(\delta)
\end{array}\right)
\end{equation}
where the interaction submatrix
\begin{equation}
W=\left(\begin{array}{cccc}
 \frac{-4}{25}{\sqrt{\frac{2}{3}}}&0&0&\frac{4}{25}{\sqrt{\frac{2}{3}}} \\
 \frac{8}{25}{\sqrt{\frac{2}{3}}}&\frac{4}{75}&\frac{-4}{75}&\frac{-8}{25}{\sqrt{\frac{2}{3}}}
\end{array}\right)
\end{equation}
and $U_3(R)=e^2\langle39d||r||41p\rangle\langle39d||r||37f\rangle/R^3=2940$ MHz $\mu$m$^3/R^3$.
  For $\delta=0$, the eigenvalues of $H$ corresponding to the $d+d$ states are $(\pm{4{\sqrt{31 \pm {\sqrt{937}}}}/75})U_3(R)$.
Two of the eigenvalues are quite small ($\pm0.033U_3(R)$), leading to poor Rydberg blockade.  The potential curves for
$\delta=-15 $ MHz are shown in Figure~\ref{fig:finestructure}.  The nearly flat potential shows that while there are no \foster
zeros for this case, the Rydberg blockade is still strongly suppressed at large distances even though the energy defect is very
small.  Even in the presence of fine structure, the blockade is still poor if $l_1<l_0$.

\section{Conclusions}

The primary result of this  paper is that the very long-range $C_3/R^3$ interactions produced by resonances $nl+nl\rightarrow
n'l_1+n''l_2$ have states with $C_3=0$ unless $l_1=l_2=l+1$.  This strongly reduces the number of possibilities for attaining
high fidelity dipole-blockade in mesoscopic atom samples.  One solution is to rely instead on quadrupole-quadrupole ($1/R^5$) or
second-order dipole-dipole (1/$R^6$) interactions to acheive blockade at high values of $n$.  As pointed out recently by the
Connecticut group
\cite{Boisseau02,Tong04} the interactions under these conditions can be quite strong.

Another possibility for attaining strong dipole blockade is to tune  $l_1=l_2=l+1$ resonances with an electric field.  For
example, the great sensitivity of $f$ states to electric fields can tune $d+d\rightarrow f+f$ into resonance at modest fields. 
At resonance, the long range potentials for this case are $\pm0.336U_3,\pm0.227U_3,\pm0.158 U_3$ and thus should lead to quite
strong Rydberg blockade at long range.

\ack

The author gratefully recognizes very helpful conversations with other members of the Wisconsin AMO Physics
group, and with C. Goebel and R. Cote.  This work was supported by the National Science Foundation, NASA, and the Army Research
Office.

\appendix
\section*{Appendix}
\setcounter{section}{1}

\noindent {\it Calculation of dipole-dipole matrix elements}

Let $l$, $l_1$, and $l_2\ge l_1$ be the angular momenta involved in the 
\foster process
\begin{equation}
nl+nl\rightarrow n'l_1+n''l_2
\end{equation}
  Since the dipole-dipole interaction
\begin{equation}
V_{DD}={\sqrt{6}e^2\over R^3}\sum_{p}C^{20}_{1p1\bar{p}}r_{Ap}r_{B\bar{p}},
\end{equation}
expressed in a coordinate system with $z$ aligned with $\bf R$,  is symmetric on atom interchange ($A\leftrightarrow B$), the
interchange-symmetric state
\begin{equation}
\ketb{lmlm'}\equiv{\ket{(lm)_A(lm')_B}+\ket{(lm')_A(lm)_B}\over\sqrt{2(1+\delta_{mm'})}}
\end{equation}mixes only with the symmetric combinations
\begin{equation}
\ketb{l_0m_0\;l_2m_2}\equiv\ket{l_0m_0l_2m_2+l_2m_2l_0m_0}/\sqrt{2}
\end{equation}  The matrix element of $V_{DD}$ is therefore
\begin{equation}
\fl \brab{lmlm'}V_{DD}\ketb{l_1m_1l_2m_2}={\bra{lmlm'}V_{DD}\ket{l_1m_1l_2m_2}+(m\leftrightarrow m')
\over\sqrt{1+\delta_{mm'}}}
\end{equation}
We use the Wigner-Eckart theorem  to write this explicitly as
\begin{equation}
\fl \brab{lmlm'}V_{DD}\ketb{l_1m_1l_2m_2}=U_3(R)
\sum_pC^{20}_{1p1\bar{p}}{C^{lm}_{l_1m_1 1p}C^{lm'}_{l_2m_2 1\bar{p}}
+(2\leftrightarrow 1)\over(2l+1)\sqrt{(1+\delta_{mm'})/6} }
\end{equation}
where 
\begin{equation}
U_3(R)={e^2\bra{nl}|r|\ket{n'l_1}\bra{nl}|r|\ket{n''l_2}\over R^3}
\end{equation}
and the reduced matrix elements are given in terms of  radial integrals of $r$ as
\begin{equation}
\bra{nl}|r|\ket{n'l_1}=\sqrt{2l_1+1}C_{l_1010}^{l0}\int_0^\infty  r P_{nl}(r)P_{n'l_1}(r)dr
\end{equation}
For the calculation of matrix elements in this paper, we have used the l-dependent core potentials of Marinescu {\it et. al}
\cite{Marinescu94} and obtained the radial wavefunctions by Numerov integration of the Schr\"{o}dinger equation.  Energy levels
were calculated using the recent quantum defect determinations of Li {\it et al.}\cite{Li03}.

\noindent {\it Reflection symmetry} 

The dipole-dipole matrix
element $\brab{l_0m_0\;l_0\bar{m}_0}V_{DD}\ketb{l_1m\;l_2\bar{m}}$is proportional to
\begin{eqnarray}
\fl\sum_p\left(C^{l_0m_0}_{l_1m 1p}C^{20}_{1p1\bar{p}}C^{l_0\bar{m}_0}_{l_2\bar{m} 1\bar{p}}
+C^{l_0m_0}_{l_2\bar{m} 1p}C^{20}_{1p\bar{p}}C^{l_0\bar{m}_0}_{l_1m 1\bar{p}}\right)\nonumber\\
=\sum_p\left(C^{l_0m_0}_{l_1m 1p}C^{20}_{1p1\bar{p}}C^{l_0{m}_0}_{l_2{m} 1{p}}
+C^{l_0{m}_0}_{l_1\bar{m} 1{p}}C^{20}_{1p\bar{p}}C^{l_0m_0}_{l_2\bar{m} 1p}\right)\label{cgsum}\end{eqnarray}
where the Clebsch-Gordan symmetry $C_{a\alpha b\beta}^{c\gamma}=-1^{a+b-c}C_{a\bar{\alpha}
b\bar{\beta}}^{c\bar{\gamma}}$ has been used along with the specific parities $(-1)^{l_1}=(-1)^{l_2}=-(-1)^{l_0}$. 
Inspection of Equation~\ref{cgsum} then shows that
\begin{equation}
\brab{l_0m_0\;l_0\bar{m}_0}V_{DD}\ketb{l_1m\;l_2\bar{m}}=\brab{l_0m_0\;l_0\bar{m}_0}V_{DD}\ketb{l_1\bar{m}\;l_2m}
\end{equation}

\section*{References}


\newcommand{\noopsort}[1]{} \newcommand{\printfirst}[2]{#1}
  \newcommand{\singleletter}[1]{#1} \newcommand{\switchargs}[2]{#2#1}

\end{document}